\documentclass{article}

\usepackage{PRIMEarxiv}

\usepackage[utf8]{inputenc} % allow utf-8 input
\usepackage[T1]{fontenc}    % use 8-bit T1 fonts
\usepackage{hyperref}       % hyperlinks
\usepackage{url}            % simple URL typesetting
\usepackage{booktabs}       % professional-quality tables
\usepackage{amsfonts}       % blackboard math symbols
\usepackage{nicefrac}       % compact symbols for 1/2, etc.
\usepackage{microtype}      % microtypography
\usepackage{lipsum}
\usepackage{fancyhdr}       % header
\usepackage{graphicx}       % graphics

\usepackage{booktabs} 

\usepackage{multirow}
\usepackage[skins]{tcolorbox}

\usepackage{subcaption}
\usepackage{xspace}
\usepackage{graphicx}
\usepackage{array}
\usepackage{xcolor}
\usepackage{bm}
\usepackage{amsmath}
\usepackage{hyperref}

\usepackage{float}
\usepackage{adjustbox}
\usepackage{listings}
\usepackage{color}
 
\lstdefinestyle{lists}{
    captionpos=b,
    abovecaptionskip=5pt,
    breaklines, 
    frame=single, 
    basicstyle=\ttfamily\tiny 
}
\title{PathOCL: Path-Based Prompt Augmentation for OCL Generation with GPT-4}

\author{
  Seif Abukhalaf \\
  Software and Emerging Technologies Lab (SAET) \\ Polytechnique Montréal \\
  Montréal, Canada \\
  \texttt{seif.abukhalaf@polymtl.ca} \\
   \And
  Mohammad Hamdaqa \\
  Software and Emerging Technologies Lab (SAET) \\ Polytechnique Montréal \\
  Montréal, Canada \\
  \texttt{mhamdaqa@polymtl.ca} \\
   \And
  Foutse Khomh \\
  SoftWare Analytics and Technologies Lab (SWAT) \\ Polytechnique Montréal \\
  Montréal, Canada \\
  \texttt{foutse.khomh@polymtl.ca} \\
}

\begin{document}
\maketitle

\begin{abstract}
The rapid progress of AI-powered programming assistants, such as GitHub Copilot, has facilitated the development of software applications. These assistants rely on large language models (LLMs), which are foundation models (FMs) that support a wide range of tasks related to understanding and generating language. LLMs have demonstrated their ability to express UML model specifications using formal languages like the Object Constraint Language (OCL). However, the context size of the prompt is limited by the number of tokens an LLM can process. This limitation becomes significant as the size of UML class models increases. In this study, we introduce PathOCL, a novel path-based prompt augmentation technique designed to facilitate OCL generation. PathOCL addresses the limitations of LLMs, specifically their token processing limit and the challenges posed by large UML class models. PathOCL is based on the concept of chunking, which selectively augments the prompts with a subset of UML classes relevant to the English specification. Our findings demonstrate that PathOCL, compared to augmenting the complete UML class model (UML-Augmentation), generates a higher number of valid and correct OCL constraints using the GPT-4 model. Moreover, the average prompt size crafted using PathOCL significantly decreases when scaling the size of the UML class models.
\end{abstract}

% keywords can be removed

\section{Introduction}
The widespread use of artificial intelligence (AI)-powered programming assistants, such as GitHub Copilot \cite{github-copilot} and ChatGPT, has introduced a paradigm shift in the ways we build software applications \cite{fan2023largeandse}. At the heart of their intelligence lie the foundation models (FMs) - AI models designed to support a wide spectrum of tasks in different modalities, empowering the modern-day chatbots and generative AI \cite{Zhou2023PFM}.

Large language models (LLMs) are one type of FMs notable for their ability to achieve general-purpose language understanding and generation. LLMs have enabled AI assistants to understand and interpret our prompts described in natural language (e.g., English). Such emergent properties have attracted the interest of software practitioners to investigate LLM applications in facilitating model-based software development tasks \cite{Abukhalaf2023, Camara2023GPTModel, Chaaben2023PromptModeling}.

One challenging task that software practitioners often encounter is expressing model specifications using formal languages such as the Object Constraint Language (OCL). Writing OCL constraints can be challenging and not straightforward, especially for novice practitioners, due to the unfamiliar syntax of the language itself \cite{Wahler2008UsingPT, bajwa2010ocl}. Therefore, manually expressing specifications becomes time-consuming when involving large UML class models and complex specifications with many entities that demand careful attention by the software practitioner to determine the proper UML classes and relations. In addition, the fact that a specification can be expressed in different OCL constraints with equivalent semantics may present further challenges for software practitioners when choosing the optimal OCL constraint \cite{Cabot2007Transformation, Lu2019AutomatedRefactoring}. Therefore, LLMs have the potential to facilitate model development by assisting software practitioners in implementing, validating, and reviewing their OCL constraints \cite{Abukhalaf2023, call_comm, Camara2023GPTModel}.
 
Different prompting techniques have been proposed to employ LLMs for UML and OCL modeling \cite{Camara2023GPTModel}. In our previous study \cite{Abukhalaf2023}, we demonstrated that relying solely on English specifications in the prompt is insufficient for generating reliable OCL constraints, and augmenting the prompt with the complete UML class model as context significantly impacts the results. However, current LLMs can only process a certain number of tokens. This limitation becomes significant when the size of the UML class models grows large enough to exceed the limited context window of the LLM.

To overcome this limitation, we draw inspiration from human problem-solving techniques for tackling complex problems. Experienced software practitioners often break down UML class models into sub-models. This approach narrows the scope, allowing them to effectively analyze different classes and semantically map elements from the English specification to their equivalent OCL expressions.

In this research, we propose \textbf{PathOCL: a path-based prompt augmentation technique for OCL generation}. PathOCL is a prompting technique that adapts the concept of chunking. It reduces the context of the prompt by selectively augmenting the subset of UML classes relevant to the English specification. 

\begin{figure*}[t]
    \centering
    \includegraphics[width=\textwidth]{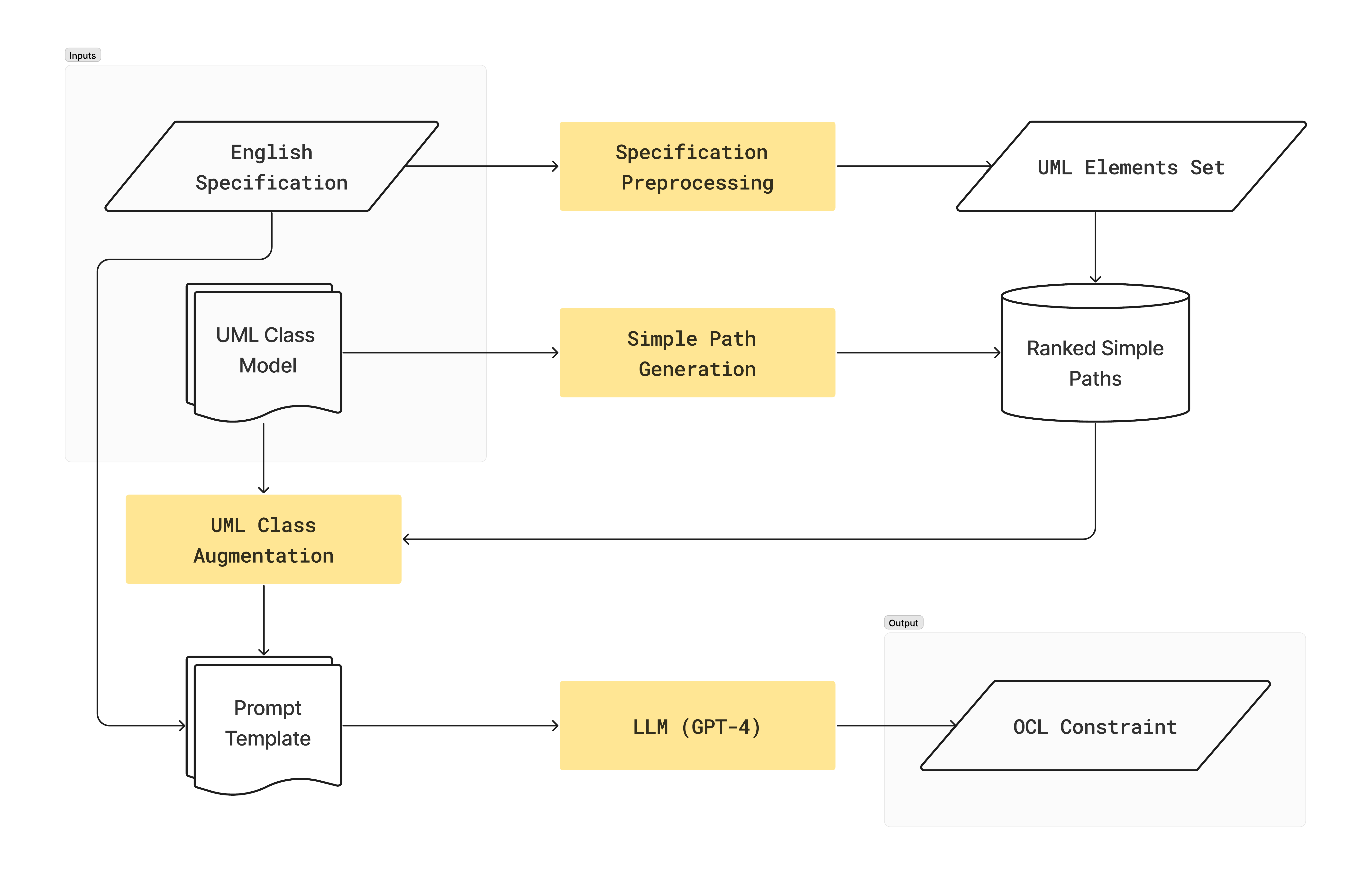}
    \caption{Overview of the PathOCL prompting process.}
    \label{fig:method_framework}
\end{figure*}

We designed PathOCL as a three-step process, illustrated in Figure \ref{fig:method_framework}. In the first step, we preprocess the English specification to extract the UML elements. In the second step, we represent the UML class models as UML graphs and generate a set of simple paths \cite{Ammann_Offutt_2018, Li2012PathTesting} that cover all the UML classes in the UML graph. We use Jaccard \cite{Jaccard1912Smilarity} and cosine similarity to rank the simple paths based on the likelihood that the set of extracted UML elements exists in the UML classes along the chosen path. In the final stage, we craft prompts augmented with the subset of UML classes based on the simple path and generate OCL constraints using the GPT-4 model. In this study, we aim to address the following research questions:

\begin{itemize}
    \item \textbf{RQ1. How effective is PathOCL prompting technique in generating valid and correct OCL constraints?}
\end{itemize}
Our study aims to empirically evaluate the effectiveness of providing a selective subset of UML classes as context (PathOCL), compared to augmenting the entire UML model (UML-Augmentation) \cite{Abukhalaf2023}. We propose the following hypotheses based on two evaluation metrics: validity and correctness scores.

\textbf{Validity.}
The initial null hypothesis (\textbf{H0-Validity}) states that there is no improvement in the validity scores when comparing the PathOCL and UML-Augmentation techniques. In contrast, the alternative hypothesis (\textbf{H1-Validity}) states that there is an improvement in the validity scores between the two techniques.

\textbf{Correctness.}
The second null hypothesis (\textbf{H0-Corre\-ctn\-ess)} states that there is no improvement in the correctness scores between the PathOCL and UML-Augmentation techniques. In the alternative hypothesis (\textbf{H1-Correctness}), we state that there is an improvement in the correctness scores between the two techniques.

\begin{itemize}
    \item \textbf{RQ2. How does the inference cost compare between PathOCL and UML-Augmentation prompting te\-chniq\-ues?}
\end{itemize}
In this research question, we assess the inference costs associated with the prompts used in both the PathOCL and UML-Augmentation techniques. The costs are calculated based on the pricing model provided by OpenAI for their GPT-4 model \footnote{https://openai.com/pricing}. In addition, we analyze the scalability of the prompts crafted using both techniques for different UML class models, including small, medium, and large-sized models.

The remaining sections of this paper are organized as follows: Section \ref{SectionII} provides the necessary background and terminology. Section \ref{SectionIII} describes our methodology. Section \ref{SectionIV} presents the empirical evaluation of PathOCL. Section \ref{SectionV} discusses our findings. Section \ref{SectionVI} introduces related works. Section \ref{SectionVII} outlines the threats to the study, and Section \ref{SectionVIII} concludes the paper.

% Overview of PathOCL framework

\section{Background and Terminology}\label{SectionII}
\textbf{Object Constraint Language (OCL)} is a formal language used to specify rules on the structure and behavior of MOF models, including UML models \cite{OCLspecs}. OCL is a declarative language; when an OCL constraint is evaluated, it does not impact the state of the model. OCL expressions can be invariants, which are applied to the attributes and associations of UML classes and must be evaluated to true for all instances of the UML model. Other types of OCL expressions can be applied to the operations of UML classes, known as pre-conditions and post-conditions.

In the context of model-driven development (MDD), OCL constraints play a crucial role in ensuring formalism and consistency during software development \cite{CabotOCLGuide}. The OCL meta-model defines the abstract syntax and semantics of OCL expressions. By formalizing constraints using the OCL meta-model, tools like the UML-based Specifications Environment (USE) \cite{GOGOLLA200727USE} can automatically verify if a specified OCL constraint conforms to the UML class model and can provide feedback to the user regarding rule violations or errors. The OCL abstract syntax defines the grammar and structure of the language, including OCL types and expressions. OCL types include data types, collection types, and message types. OCL expressions can also include property expressions, if expressions, iterator expressions, variable expressions, and other expressions.

\textbf{Prompt engineering} is a systematic approach in designing prompts that effectively generate a response from LLMs. The design and implementation of prompts have a critical impact on the performance of LLMs, and several prompting techniques have been proposed and proven to be effective \cite{liu2021pretrainprompt}. One commonly used type of prompt is zero-shot prompting, where the prompt context does not include examples for the LLM to follow. In contrast, few-shot prompting involves providing input-output pair examples in the context to guide and customize the response format of the LLM. Advanced reasoning techniques, such as Chain-of-Thoughts (CoT) \cite{wei2023chainofthought}, where the LLM is prompted through step-by-step reasoning, have also been used to improve problem-solving abilities in logical tasks, critical thinking, and complex programming \cite{le2023codechain}. Another prompting technique involves retrieving chunks of relevant documents based on the given query and incorporating them as context, which plays a pivotal role in reducing hallucinations by grounding the LLM on the augmented information \cite{ram2023RALM}.

\textbf{Graphs} are fundamental data structures used to represent relationships between entities. A graph consists of nodes (vertices) and edges that connect pairs of nodes. Formally \cite{Ammann_Offutt_2018}, a graph $G$ is:

\begin{itemize}
    \item A set of \textit{N} \textit{nodes}, where \textit{N} $\ne$ $\phi$.
    \item A set of $N_0$ \textit{initial nodes}, where $N_0$ $\subseteq$ \textit{N} and $N_0$ $\ne$ $\phi$.
    \item A set of $N_f$ \textit{final nodes}, where $N_f$ $\subseteq$ \textit{N} and $N_f$ $\ne$ $\phi$.
    \item A set of \textit{E} \textit{edges}, where \textit{E} is a subset of \textit{N} $\times$ \textit{N}.
\end{itemize}

A graph is directed when edges have a defined direction from one node $n_i$ to another $n_j$, denoted as ($n_i$, $n_j$). A path is a sequence [$n_1$, $n_2$, $...$, $n_M$] of nodes, where each pair of adjacent nodes, ($n_i$, $n_i+1$), 1 $\leq$ \textit{i} $\leq$ \textit{M}, is in the set E of edges.

\textbf{Simple paths} \cite{Ammann_Offutt_2018} refer to sequences of edges from node $n_i$ to node $n_j$ where no node is repeated. In other words, a simple path is a path through a graph that does not revisit any node.

\section{Methodology}\label{SectionIII}
In this section, we present PathOCL, our approach for generating OCL constraints given a UML class model and the natural language specification in English. The primary objective of PathOCL is to overcome the limitations in the number of tokens an LLM can process, which can occur when the size of the UML class model exceeds the context size of the prompt. PathOCL attempts to emulate the process that an experienced software practitioner follows when formulating OCL constraints. The expert typically starts by reducing the UML class model into a subset of classes relevant to the context of the given English specification.

PathOCL requires two inputs: (a) the OCL specification written in natural language (i.e., English) and (b) the UML class model. Figure \ref{fig:method_framework} provides an overview of the PathOCL prompting approach. PathOCL consists of three main steps: (a) English specification preprocessing, (b) simple paths generation and ranking, and (c) prompt augmentation with the selective subset of UML classes to generate OCL constraint. We present the airport domain model as a running example to motivate PathOCL and explain each step of the approach in detail.

% Airpot Motivating Example
\begin{figure}[!htbp]
    \centering
    \includegraphics[width=0.4\textwidth]{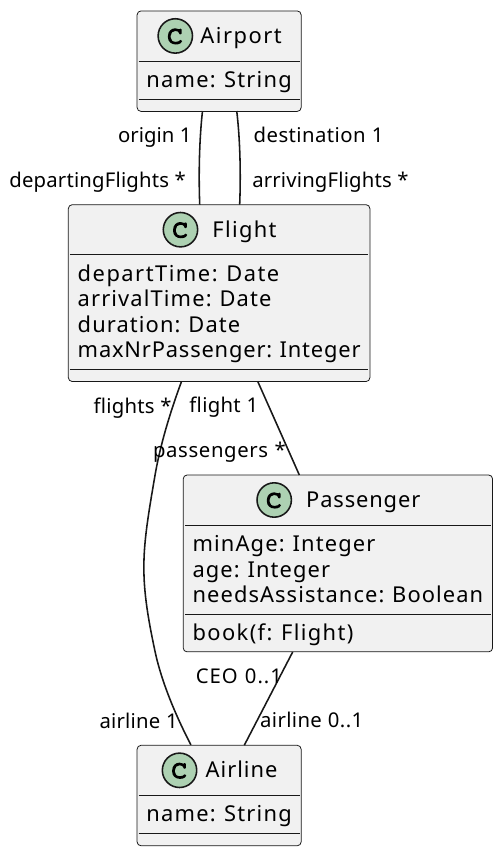}
    \caption{The UML class diagram of the airport domain model.}
    \label{fig:uml-airport}
\end{figure}

\subsection{Case Study}
The airport domain model, shown in Figure \ref{fig:uml-airport}, represents the organization and functionality of a fictional airline system. The airport class plays a vital role as departure and arrival points for flights. The flight class, with a specific number of passengers, originates from an airport and travels to designated airports. The airline class is responsible for operating and supervising these flights. It is common for airlines to have a Chief Executive Officer (CEO) who maintains a close affiliation with the company. We will run the example by introducing the following English specification: 

% Demo specification
\begin{center}
    \textit{“The maximum number of passengers on any flight may not exceed 1000.”}
\end{center}

\subsection{Specification Preprocessing}
The first step in PathOCL is to extract UML elements from the provided English specification. These elements serve as the heuristic to rank the set of simple paths generated from the UML graph. Specification preprocessing consists of three steps: (a) tokenization, (b) part-of-speech (PoS) tagging, and (c) lemmatization. We use spaCy trained English pipeline “en\_core\_web\_sm”, which includes the required components for extracting the UML elements. spaCy is widely adopted for its robust pre-trained statistical models that deliver exceptional performance and reliability in natural language processing (NLP) \cite{Honnibal2020spaCy}.

In the previous study conducted by Salemi et al. \cite{Salemi2016OCL}, the authors established mapping rules to extract UML elements from their defined English metamodel. As presented in Table \ref{tab:mapping-rules}, we use the same rules to identify the UML elements from the English specification.

% ========================================================================
% Mapping rules from Salemi et al.
\begin{table*}[b]
\centering
\caption{Mapping rules between the UML elements from the English specification to the equivalent OCL expression.}
\label{tab:mapping-rules}
\resizebox{\textwidth}{!}{
\begin{tabular}{@{}lllcl@{}}
\toprule
\textbf{Rule} & \textbf{English Element}                   & \textbf{English Example} & \multicolumn{1}{l}{\textbf{OCL Element}}    & \textbf{OCL Example} \\ \midrule
Rule 2        & \textit{IsPropertyOfSign}                  & Name of customer         & \multirow{4}{*}{\textit{AttributeCallExp}}  & Customer.name        \\
Rule 4        & \textit{PrefixElement}                     & Valid CustomerCard       &                                             & CustomerCard.valid   \\
Rule 6        & \textit{TransitiveVerb}        & Customer has names       &                                             & Customer.name        \\
Rule 7        & \textit{PrepositionConjunction}            & Transaction with points  &                                             & Transaction.point    \\ \midrule
Rule 3        & \textit{PossessiveDeterminer}              & Customer’s cards         & \multirow{2}{*}{\textit{NavigationCallExp}} & Customer.cards       \\
Rule 10       & \textit{Noun  (mapped to attribute/roles)} & CustomerCard owner       &                                             & CustomerCard.owner   \\ \midrule
Rule 11       & \textit{Noun (mapped to Class)}            & Service                  & \textit{UMLElement}                         & Service              \\ \bottomrule
\end{tabular}}
\end{table*}

According to their English metamodel, \textit{IsPropertyOfSign} is a sub-phrase that links two semantic elements using “of”. \textit{PrefixElement} is a semantic element placed immediately before another semantic element. \textit{TransitiveVerb} is a verb that takes one or more objects. \textit{PrepositionConjunction} is a preposition, such as “in” and “with”, describing a relationship between two sub-phrases in a sentence. \textit{PossessiveDeterminer} is a sub-phrase of determiners that modify a noun by attributing possession to someone or something. The examples provided in Table \ref{tab:mapping-rules} are applied to the Royal \& Loyal domain model presented in Figure \ref{figure:royalandloyal}. 

When we applied spaCy to English specifications, we observed that the POS tagger identifies the English elements from Table \ref{tab:mapping-rules} as nouns and adjectives. Therefore, we extract the nouns and adjectives from the English specification and add them to the UML elements set. Continuing with the running example, we obtain the following set of UML elements:

% Extracted elements
\begin{center}
    \textit{Nouns} $\to$ \{number, passengers, flight\}
    \\
    \textit{Adjective} $\to$ \{maximum\}
\end{center}

In the final step, we employ spaCy for lemmatization. This process transforms the extracted UML elements into their base form (lemma) by removing inflectional endings, such as the “-ing” suffix \cite{Honnibal2020spaCy}. We store the UML elements in sets, which allows us to eliminate any duplicates. These sets form the basis for ranking the simple paths. After processing the previously obtained set, we get the following result:

% Processed elements
\begin{center}
    \textit{UML elements} $\to$ \{number, passenger, flight, maximum\}
\end{center}

\subsection{Simple Path Generation}
In our previous study, we introduced a prompting technique, UML-Augmentation, that augments the prompt with the complete UML class model to generate OCL constraints \cite{Abukhalaf2023}. However, this technique encounters limitations when the UML class model size exceeds the context size that the LLM can process.

We address this limitation by decomposing the UML class model into sub-models. To achieve this, we represent the UML class models as graphs to cover the simple paths in the equivalent UML graph. The UML graph serves as an abstract representation of the UML class diagram. The relationship between two UML classes can be directional, resulting in a directed graph representation of the UML graph. In this representation, each node denotes a UML class, and each edge represents the relationship between two UML classes.

In PathOCL, we generate simple paths that cover all the UML classes in the UML graph. This is due to the possibility that an English specification can have multiple semantically equivalent OCL constraints expressed differently \cite{Lu2019AutomatedRefactoring}. To accomplish this, we apply the brute force solution proposed by Li et al. \cite{Li2012PathTesting} to cover all the simple paths for a given graph. Their solution consists of three algorithms that take the graph and the set of initial and final nodes as input to generate the simple paths. 

In the current setting of PathOCL, we include all the UML classes in the initial and final input sets. As a result, we obtain a set of simple paths that cover all UML classes in the UML graph. In addition, we also include the UML classes as individual nodes in the set, as it is possible to express an OCL constraint using just a single UML class, without the need to navigate through associations.

Applying the solution to the airport domain model in Figure \ref{fig:uml-airport}, we generate a set of simple paths that cover all UML classes, in addition to the individual UML classes. A subset of the simple paths is shown in Listing \ref{listing:simple-paths}.

% Simple paths listing
\begin{lstlisting}[caption={The subset of simple paths generated for the airport domain model.}, label={listing:simple-paths}, captionpos=b, float=b]
[Airport]
[Flight]
[Passenger]
[Airline]
...
[Airline,Flight]
[Passenger,Airline]
[Airline,Passenger]
...
[Passenger,Airline,Flight,Airport]
[Airline,Passenger,Flight,Airport]
[Passenger,Airline,Flight,Passenger]
\end{lstlisting}

\subsection{Simple Path Ranking}
Covering all simple paths in a UML graph would result in a significant number of prompts for generating OCL constraints. To manage this, we rank the simple paths based on the likelihood that the English specification applies to the subset of UML classes within each simple path. We then choose the top-k prompts for generating OCL constraints.

We achieve this by assigning each simple path a similarity score. This score measures the textual similarity between two sets of words: (a) the UML elements extracted from the English specification, and (b) the properties of the UML classes in the simple path. Note that we only consider the name of the UML class properties, not their datatype or cardinality.

Similar to the UML elements set, we apply lemmatization to the UML class properties to obtain the same base form of words. Furthermore, we only choose the roles at the end of the next UML class in the simple path. For instance, in the simple path “[Airline, Flight]” from Listing \ref{listing:simple-paths}, we only consider the role “flights” between the “Airline” and “Flight” classes. For a single UML class, such as “[Airport]”, we only include its properties. Storing the UML properties as sets allows us to remove any duplicates that might occur when the initial UML class appears as the final node, as in the last simple path in Listing \ref{listing:simple-paths}. The following UML properties set is extracted for the simple path “[Airline, Flight]”: 

% UML properties set
\begin{center}
    \textit{UML Properties Set} $\to$ \{Airline, Flight, name, flight, departtime, arrivaltime, duration, maxnrpassenger\}
\end{center}

Now, we use two similarity metrics to assign scores for each simple path: (a) Jaccard similarity and (b) cosine similarity.

\subsubsection{\textbf{Jaccard Similarity}}
The Jaccard index is a useful and efficient method for measuring the overlap of elements in two sets \cite{Jaccard1912Smilarity}. It is computed using the given equation:

\begin{equation}
\label{(1)}
Jaccard(E,P) = \frac{|E \cap P|}{|E \cup P|}
\end{equation}

Here, \textit{E} represents the set of UML elements extracted from the English specification, and \textit{P} represents the set of properties for the UML classes in the simple path. It is important to note that the Jaccard similarity score is based on the exact matching between the sets of words and does not take into account synonyms and word context. This implies that differences in words, such as abbreviations (e.g., serviceNr), are not accounted for.

When Jaccard similarity is applied to the generated simple paths in Listing \ref{listing:simple-paths}, the outcome is the ranked list shown in Listing \ref{listing:jaccard-ranks}. As anticipated, the top-ranked simple paths include the UML classes relevant to the English specification.

% Jaccard Ranked Simple paths listing
\begin{lstlisting}[caption={The ranked subset of simple paths obtained by applying Jaccard similarity.}, label={listing:jaccard-ranks}, captionpos=b, float=t]
['Flight', 'Passenger'] -> 0.25
['Flight'] -> 0.12
...
['Passenger', 'Airline', 'Passenger'] -> 0.0
['Airport'] -> 0.0
\end{lstlisting}
% 

% Cosine Ranked Simple paths listing
\begin{lstlisting}[caption={The ranked subset of simple paths obtained by applying cosine similarity.}, label={listing:cosine-ranks}, captionpos=b, float=t]
['Airline', 'Passenger'] -> 0.40
['Airline', 'Flight', 'Passenger'] -> 0.37
...
['Flight'] -> 0.31
['Passenger'] -> 0.30
\end{lstlisting}

\subsubsection{\textbf{Cosine Similarity}}
In certain cases, exact matching may not be the most effective method to rank simple paths. This issue becomes clear when UML class elements are not labeled following a specific convention. For example, the attribute “maxNrPassenger” from Figure \ref{fig:uml-airport}, could lead to mismatches due to minor variations in the name. As a result, we expand our ranking metrics by matching word embeddings and assigning similarity scores using the cosine similarity metric. The cosine score is measured using the given equation:

% Cosine metric
\begin{equation}
\label{(2)}
Cosine(e, p) = \frac{e \cdot p}{|e||p|}
\end{equation}

Here, \textit{e} represents the embedding vector of one element from the UML elements set \textit{E}, and \textit{p} represents the embedding vector of one property from the UML class properties set \textit{P}. We compute the cosine similarity between the vectors of each element, resulting in a matrix for all possible pairs in the two sets. We take the average of the matrix and assign it as the similarity score for the corresponding simple path.

We also apply cosine similarity to the same set of simple paths generated from the airport domain model in Listing \ref{listing:simple-paths}. As shown in Listing \ref{listing:cosine-ranks}, the top-k simple paths and their ranking scores are different from the results obtained using Jaccard similarity. This is attributed to the fact that word embeddings incorporate the semantics of the words, enabling them to take into account synonyms and variations.

\subsection{OCL Constraint Generation}
PathOCL is versatile and can be used with any LLM. In this study, we consider the GPT-4 model by OpenAI. The behavior of GPT-4 models can be guided by setting instructions as the system prompts \cite{gpt-docs}. Our template for the system prompt is shown in Listing \ref{listing:system-prompt}.

In addition, we have designed a fixed prompt template that serves as the user prompt for generating OCL constraints. The prompt is divided into two parts: (a) the English specification, and (b) the UML classes in the simple path. For each class in the selected path, we augment the prompt with its attributes, operations, and roles with their cardinality. This context is formatted in JSON, as shown in Listing \ref{listing:user-prompt}.

To conclude the running example, we apply the user template to the top-1 simple path from Listing \ref{listing:jaccard-ranks}. The properties of both the “Flight” and “Passenger” classes are augmented in the prompt. Below is a correct OCL implementation successfully generated by the GPT-4 model:

\begin{center}
    \textit{context Flight inv: self.passengers->size() <= 1000}
\end{center}

% System prompt listing
\begin{lstlisting}[caption={The system prompt template to instruct GPT-4.}, label={listing:system-prompt}, captionpos=b, float=t]
As a system designer with expertise in UML modeling and OCL constraints, your role is to assist the user in writing OCL constraints. The user will provide you with the following information: 

(1) The specification in natural language.
(2) The UML classes and their properties (attributes, operations, associations). 

Your objective is to generate a valid OCL constraint according to the provided UML classes. Please do not provide explanation. Put your solution in a <OCL> tag.
\end{lstlisting}
% 

% User prompt template listing
\begin{lstlisting}[caption={The user prompt template to generate OCL constraints.}, label={listing:user-prompt}, captionpos=b, float=t]
-- OCL specification
<English specification>

-- UML properties of class <class name>
{
    "attributes":
    [
        {
        <attribute name>: <data type>
        }
    ],
    "operations":
    ]
        {
            <operation name>: <data type>
        }
    ],
    "associations": 
    [
        {
            "target": <associated class>,
            "role": <association name>,
            "multiplicity": <cardinality>
        }
    ]
}

-- OCL constraint
\end{lstlisting}

\section{Empirical Evaluation}\label{SectionIV}
% Royal and Loyal model diagram
\begin{figure}
    \centering
    \includegraphics[width=0.8\textwidth]{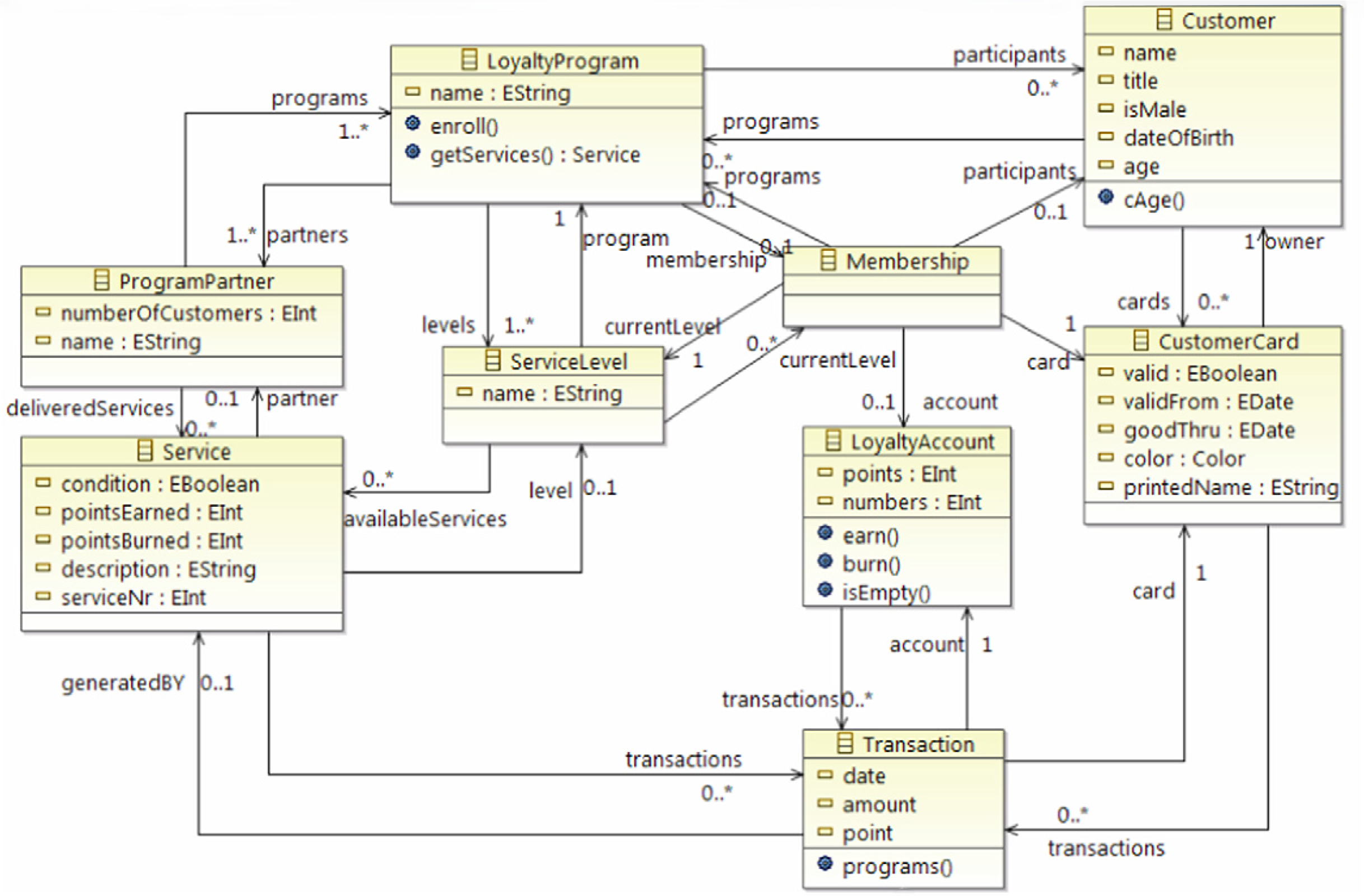}
    \caption{The UML class diagram of the Royal \& Loyal domain model.}
    \label{figure:royalandloyal}
\end{figure}
% 
% 

% Replication package.
The dataset used in our experiments is publicly available on Zenodo\footnote{https://zenodo.org/doi/10.5281/zenodo.10841785}. It contains 15 UML class models, with a total of 168 English specifications. We present the Royal \& Loyal domain model for a detailed analysis of our experiments. The Royal \& Loyal model is a hypothetical company that manages loyalty programs for businesses \cite{Kleppe_Warmer, bajwa2010ocl}. Figure \ref{figure:royalandloyal} illustrates the UML class diagram of the case study.

\subsection{\textbf{Models and Configurations}}
We use the GPT-4 model as the LLM to generate OCL constraints. This model is configured with default parameters: (a) a maximum output token limit of 256, and (b) a temperature setting of zero to enable reproducing the results \cite{gpt-docs}. To generate the word embeddings, we employ the “all-MiniLM-L6-v2” model provided by Sentence-BERT \cite{reimers-2019-sentence-bert}.

\subsection{\textbf{Evaluation Metrics}} 
\subsubsection{\textbf{Validity@K}}
An OCL constraint is considered valid if it conforms with the syntactical structure and formatting rules of OCL \cite{OCLspecs}. Therefore, the validity score (Validity) represents the percentage of valid OCL constraints generated. This score is calculated by dividing the number of OCL constraints, successfully compiled by the USE tool, by the total number of OCL constraints. The formula is as follows:

\begin{equation}
\label{(3)}
Validity = \frac{Valid \ OCL \ Constraints}{Total \ OCL \ Constraints}
\end{equation}

\subsubsection{\textbf{Correctness@K}}
A correct OCL constraint must ensure that the expression is valid and accurately implements the constraints and rules of the English specification within the context of its target UML classes. Thus, the correctness score (Ccorrectness) represents the percentage of correct OCL constraints generated. This score is calculated by dividing the number of OCL constraints deemed correct by the software modeler, using the USE tool, by the total number of OCL constraints. The formula is as follows:

\begin{equation}
\label{(4)}
Correctness = \frac{Correct \ OCL \ Constraints}{Total \ OCL \ Constraints}
\end{equation}

\subsubsection{\textbf{McNemar's Test}}
Each generated OCL constraint is evaluated to verify its validity and correctness, resulting in the OCL constraint being classified as either valid or invalid, and correct or incorrect. In our study, we used the same dataset across all experiments. Therefore, we apply McNemar's test \cite{McNemar1947McNemarsTest} to statistically test our null hypotheses and compare the PathOCL and UML-Augmentation prompting techniques. This test is commonly used for paired nominal data based on a 2x2 contingency matrix. 

\subsection{RQ1. How effective is PathOCL prompting technique in generating valid and correct OCL constraints?}
In our hypotheses, we stated that there is no improvement in both the validity and correctness scores when using PathOCL as opposed to UML-Augmentation. However, in our previous study \cite{Abukhalaf2023}, we used the UML-Augmentation prompting technique with Codex as the LLM. Therefore, our first step is to apply UML-Augmentation with the GPT-4 model, which will enable us to statistically evaluate the hypotheses.

\subsubsection{\textbf{GPT-4 vs. Codex}}
Using the UML-Augmentation prompting technique, we found that GPT-4 maintained similar scores in generating valid OCL constraints compared to Codex. However, GPT-4 showed a significant 26.5\% improvement in generating correct OCL constraints. The comparison of Codex and GPT-4 results is displayed in the top section of Table \ref{tab:all-comparison-results}.

Our findings indicate that the advancements in reasoning abilities of GPT-4 have improved its understanding of English specifications. This leads to a better alignment of OCL constraints with the semantics of UML class models. However, the syntax validity of OCL constraints remains unchanged. This implies that using the entire UML class model as context could potentially hinder the ability of the GPT models to recall the correct UML properties.

% Comparison table
\begin{table}[t]
\centering
\caption{Overall validity and correctness scores for PathOCL and UML-Augmentation prompting techniques.}
\label{tab:all-comparison-results}
\vspace{6pt}
\resizebox{0.65\textwidth}{!}{
\begin{tabular}{@{}lll@{}}
\toprule
\textbf{Prompting Techniques}      & \multicolumn{1}{c}{\textbf{Validity (\%)}} & \multicolumn{1}{c}{\textbf{Correctness (\%)}} \\ \midrule
UML-Aug. \& Zero-Shot (Codex) & 48.5                                       & 35.1                                          \\
UML-Aug. \& Zero-Shot (GPT-4) & 47.8                                       & 44.0                                          \\ \midrule
PathOCL - Cosine (GPT-4)                & \textbf{57.7}                              & \textbf{47.6}     \\                            
PathOCL - Jaccard (GPT-4)                & \textbf{61.9}                              & \textbf{46.4}                                 \\ 
\bottomrule
\end{tabular}
}
\end{table}
% 

% Top-K and Ranking table
\begin{table}[b]
\centering
\caption{Overall validity and correctness scores for PathOCL with top-k prompts.}
\label{tab:top-k-results}
\vspace{6pt}
\resizebox{0.65\textwidth}{!}{
\begin{tabular}{@{}lllllll@{}}
\toprule
\multicolumn{1}{c}{\multirow{2}{*}{\textbf{Ranking Metrics}}} & \multicolumn{3}{c}{\textbf{Validity (\%)}}                                                                    & \multicolumn{3}{c}{\textbf{Correctness (\%)}}     \\ \cmidrule(l){2-7} 
\multicolumn{1}{c}{}                                          & \multicolumn{1}{c}{\textbf{top-3}} & \multicolumn{1}{c}{\textbf{top-5}} & \multicolumn{1}{c}{\textbf{top-10}} & \textbf{top-3} & \textbf{top-5} & \textbf{top-10} \\ \midrule
Jaccard                                                       & 44.6                               & 50.0                               & \textbf{61.9}                                & 36.3           & 39.8           & \textbf{46.4}            \\ \midrule
Cosine                                                        & 46.4                               & 50.0                               & \textbf{57.7}                                & 38.0           & 40.4           & \textbf{47.6}            \\ \bottomrule
\end{tabular}
}
\end{table}

\subsubsection{\textbf{PathOCL Evaluation}}
We deem an OCL constraint as valid and correct if it is successfully generated by GPT-4 using the top-k prompt. However, covering all simple paths could lead to a substantial set of prompts to evaluate. Therefore, we limit our experiments to only consider the top-10 prompts. Our analysis begins with evaluating the syntax validity of the generated OCL constraints. As shown in Table \ref{tab:top-k-results}, we noticed a consistent increase in the number of valid OCL constraints when testing various prompts that cover different paths of UML classes, using both Jaccard and cosine ranking metrics.

% Syntax errors table
\begin{table*}[h]
\centering
\caption{Overall percentage of the syntax error categories for PathOCL (Jaccard) and UML-Augmentation prompting techniques.}
\label{tab:syntax-errors}
\resizebox{\textwidth}{!}{
\begin{tabular}{@{}lcccc@{}}
\toprule
\textbf{Prompting Method}      & \multicolumn{1}{l}{\textbf{Undefined Operation}} & \multicolumn{1}{l}{\textbf{Parsing Error}} & \multicolumn{1}{l}{\textbf{IterExp Invalid Source}} & \multicolumn{1}{l}{\textbf{Signature Mismatch}} \\ \midrule
UML Info. \& Zero-Shot & 80.5\%                                           & 15.2\%                                     & 4.1\%                                               & 0.2\%                                           \\
PathOCL (Jaccard)                & 82.5\%                                           & 9.6\%                                      & 5.7\%                                               & 2.2\%                                           \\ \bottomrule
\end{tabular}}
\end{table*}

% 

% Best performing setting
To gain insights into the main causes of GPT-4 generating invalid OCL constraints, we used the OCL compiler from the USE tool to detect and summarize the syntax errors. We found that the most common error in the invalid OCL constraints occurred when the GPT-4 model referred to undefined and incorrect properties from the UML classes. The USE tool reported this syntax error as an “undefined operation”, which can have different interpretations in certain cases.

An example is demonstrated in Listing \ref{listing:undefined-operation}, where the GPT-4 model referred to the role “programs” that is not defined as a property of the “Transaction” class. However, we observed that when GPT-4 was prompted with additional prompts of paths involving different UML classes, it successfully generated a valid OCL constraint for the same example. This aligns with the fact that an OCL constraint can have a large number of alternative and semantically equivalent implementations \cite{Cabot2007Transformation, Lu2019AutomatedRefactoring}. Therefore, exposing the GPT-4 model to different sets of UML classes allows for exploring variations of unique OCL constraints that would result in valid and correct implementations.

% Undefined Operation
\begin{lstlisting}[caption={An example of the “undefined operation” error.}, label={listing:undefined-operation}, captionpos=b, float=t]
English specification:
The owner of a customer card must participate in at least one loyalty program.

OCL constraint:
context CustomerCard
inv: self.transactions->exists(t | t.programs->notEmpty())

Syntax error:
Undefined operation named 'programs' in expression 'Transaction.programs()'.
\end{lstlisting}

Another category of errors identified is related to the structure of the OCL constraint itself. Parsing errors occur when the OCL parser of the USE tool encounters an OCL constraint that does not conform with the defined syntax by the OCL grammar and rules \cite{OCLspecs}. A closer investigation revealed that some of the OCL constraints were improperly formatted, as can be seen in Listing \ref{listing:parsing-error}, where the keyword “inv” had been used instead of “pre”, “post”, or “body”. 

% Parsing Error
\begin{lstlisting}[caption={An example of the parsing error.}, label={listing:parsing-error}, captionpos=b, float=b]
English specification:
The maximum age of participants in loyalty programs is 70.

OCL constraint:
context Transaction::programs()
inv: self.participants->forAll(p | p.age <= 70)

Syntax error:
No viable alternative at input 'inv'.
\end{lstlisting}

In addition to the errors related to the UML properties and the OCL syntax, we have identified an error with the iterator expressions, which are commonly used in OCL constraints. For example, the “select” iterator in Listing \ref{listing:iterexp-invalid-source} is being applied on a single object of type “CustomerCard” instead of a set of objects. This is attributed to the fact that the association “card” between the class “Membership” and “CustomerCard” has a cardinality of one. Another error occurs when the operation referenced in the OCL constraint does not match its signature as defined in the UML class. Listing \ref{listing:signature-mismatch} provides an example where the operation “enroll()” is referenced with an extra input parameter that does not match the definition in the “LoyaltyProgram” class.

Table \ref{tab:syntax-errors} summarizes the percentages of each error category. We chose PathOCL with the Jaccard metric as it is the best-performing setting to present the results. We can observe a marginal difference in the syntax error categories between the PathOCL and UML-Augmentation prompting techniques. This suggests that, regardless of the prompting technique used, the GPT-4 model still faces limitations in accurately referencing the UML properties according to the English specification.

% IterExp Invalid Source listing
\begin{lstlisting}[caption={An example of the “IterExp Invalid Source” error.}, label={listing:iterexp-invalid-source}, captionpos=b, float=t]
English specification:
The number of valid cards for every customer must be equal to the number of programs in which the customer participates.

OCL constraint:
context Membership
inv: self.card->select(c | c.valid = true)->size() = self.programs->size()

Syntax error:
Source of 'select' expression must be a collection, found source expression of type 'CustomerCard'.
\end{lstlisting}

% Signature Mismatch listing
\begin{lstlisting}[caption={An example of the “Signature Mismatch” error.}, label={listing:signature-mismatch}, captionpos=b, float=t]
English specification:
Every customer who enters a loyalty program must be of legal age.

OCL constraint:
context LoyaltyProgram::enroll(c: Customer)
    pre: c.age >= 18

Syntax error:
This signature of operation 'enroll' does not match its previous declaration in class 'LoyaltyProgram'.
\end{lstlisting}

\begin{figure*}
    \centering
    \includegraphics[width=0.75\textwidth]{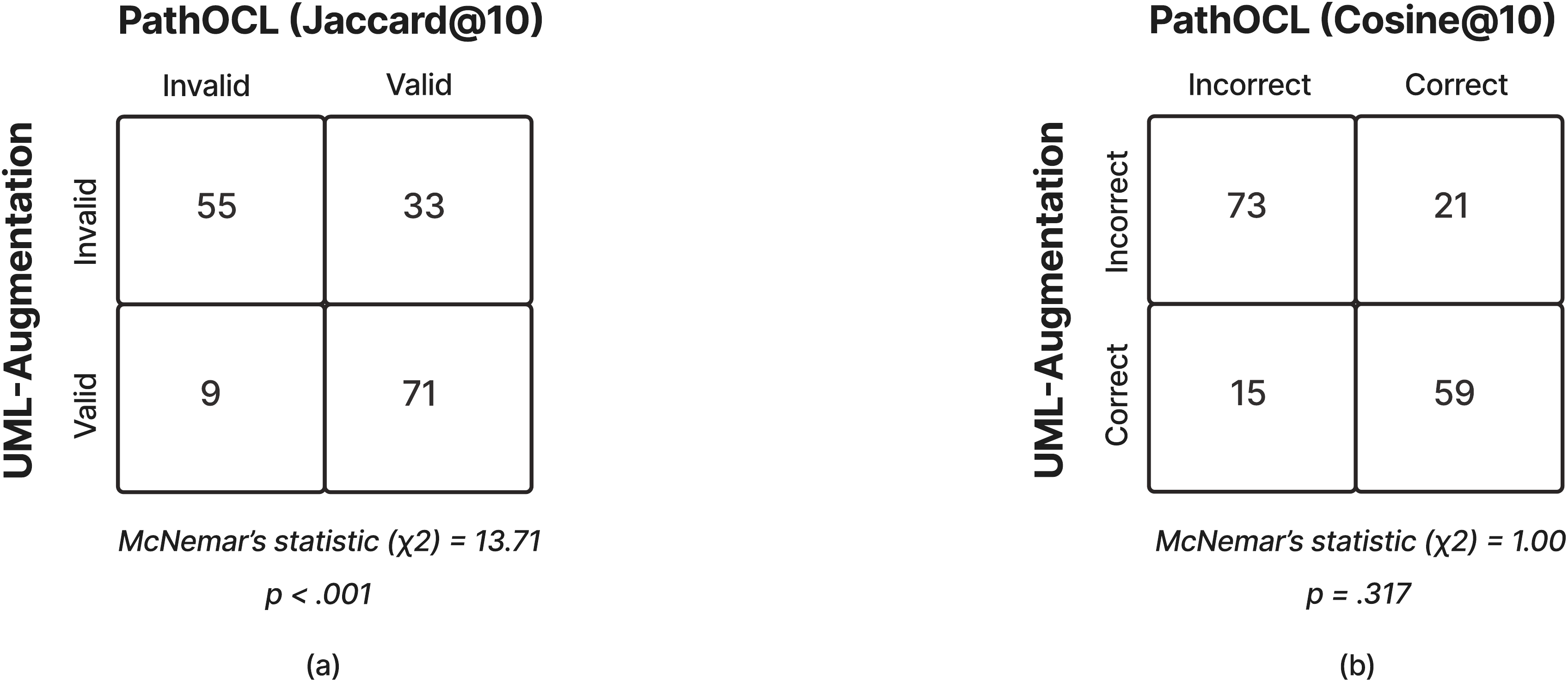}
    \caption{The McNemar's 2x2 contingency tables are used to evaluate both the PathOCL and UML-Augmentation prompting techniques. Table (a) for the validity scores and Table (b) for the correctness scores.}
    \label{fig:mcnemar-tables}
\end{figure*}

\subsubsection{\textbf{PathOCL vs. UML-Augmentation.}}
In our null hypothesis (\textbf{H0-Validity}), we stated that providing a selective subset of UML classes has no improvement in the validity score compared to augmenting the entire UML class model. However, our empirical results in Table \ref{tab:all-comparison-results} show that the PathOCL prompting technique demonstrates a substantial improvement in the validity score by 30\% using Jaccard and by 20\% using cosine when compared to the UML-augmentation technique. 

To measure the improvement statistically, we use McNemar's test to compare the validity scores of both PathOCL and UML-Augmentation techniques. McNemar's test is applied to the 2x2 contingency table (a) shown in Figure \ref{fig:mcnemar-tables}. The highlighted numbers in this table are the off-diagonal cells as they indicate the differences in generating valid OCL constraints between the prompting techniques. We use the following equation to calculate McNemar's statistic ($\chi^2$):

\begin{equation}
\label{(5)}
\chi^2 = \frac{(B-C)^2}{B+C}
\end{equation}

Where B is the second cell from the first row, and C is the first cell from the second row. We set ($\alpha$ = 0.05) as the significance threshold to match with the p-value and decide on the null hypothesis. Since McNemar's test follows the ($\chi^2$) distribution with one degree of freedom, we look up the one-tail p-value via the ($\chi^2$) distribution. For table (a), ($\chi^2$) is equal to 13.71, and therefore we obtain a p-value of 0.0002, which is below the significance threshold ($\alpha$ = 0.05). As a result, we confidently reject the null hypothesis (\textbf{H0-Validity}) and accept the alternative hypothesis (\textbf{H1-Validity}). Thus, we conclude that using the PathOCL prompting technique improves the number of valid OCL constraints generated when compared to the UML-Augmentation technique.

Similarly, we propose that there is no improvement in the correctness score (\textbf{H0-Correctness}) when using PathOCL compared to UML-Augmentation. To statistically test our null hypothesis, we also apply McNemar's test to compare the empirical results from table (b) in Figure \ref{fig:mcnemar-tables}. Using equation \ref{(5)}, we calculate McNemar's statistic ($\chi^2$) to be equal to 1.00, resulting in a p-value of .317. This value is higher than our significance threshold ($\alpha$ = 0.05), therefore, we cannot confidently reject the null hypothesis (\textbf{H0-Correctness}) and cannot accept the alternative hypothesis (\textbf{H1-Correctness}). Regardless, even if the statistical difference may not be significant, there is still a slight increase in the number of correctly generated OCL constraints, as observed in Table \ref{tab:all-comparison-results}.

\subsection{RQ2. How does the inference cost compare between PathOCL and UML-Augmentation prompting techniques?}
In this research question, we analyze the computational costs associated with using both the PathOCL and the UML-Augmentation techniques to prompt the GPT-4 model. We specifically analyze PathOCL with the Jaccard metric as it overall outperforms when compared with the cosine metric.

The inference costs are calculated based on the pricing model of OpenAI for their GPT-4 model, which charged \$0.003 per 1K input tokens consumed at the time of this study \footnote{November, 2023}. These costs only account for the prompts that generated valid and correct OCL constraints. 

\begin{figure}
    \centering
    \includegraphics[width=0.8\textwidth]{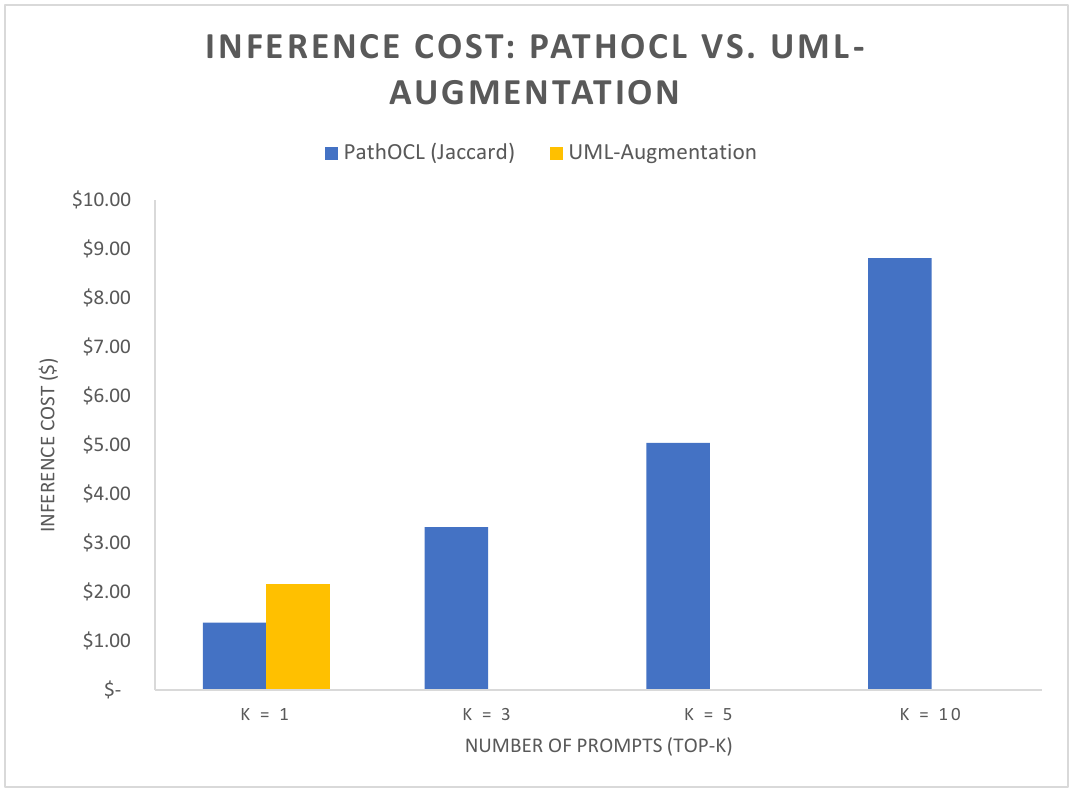}
    \caption{Comparative inference cost analysis for PathOCL (Jaccard) vs. UML-Augmentation}
    \label{fig:prompt-cost}
\end{figure}

\begin{figure}
    \centering
    \includegraphics[width=0.8\textwidth]{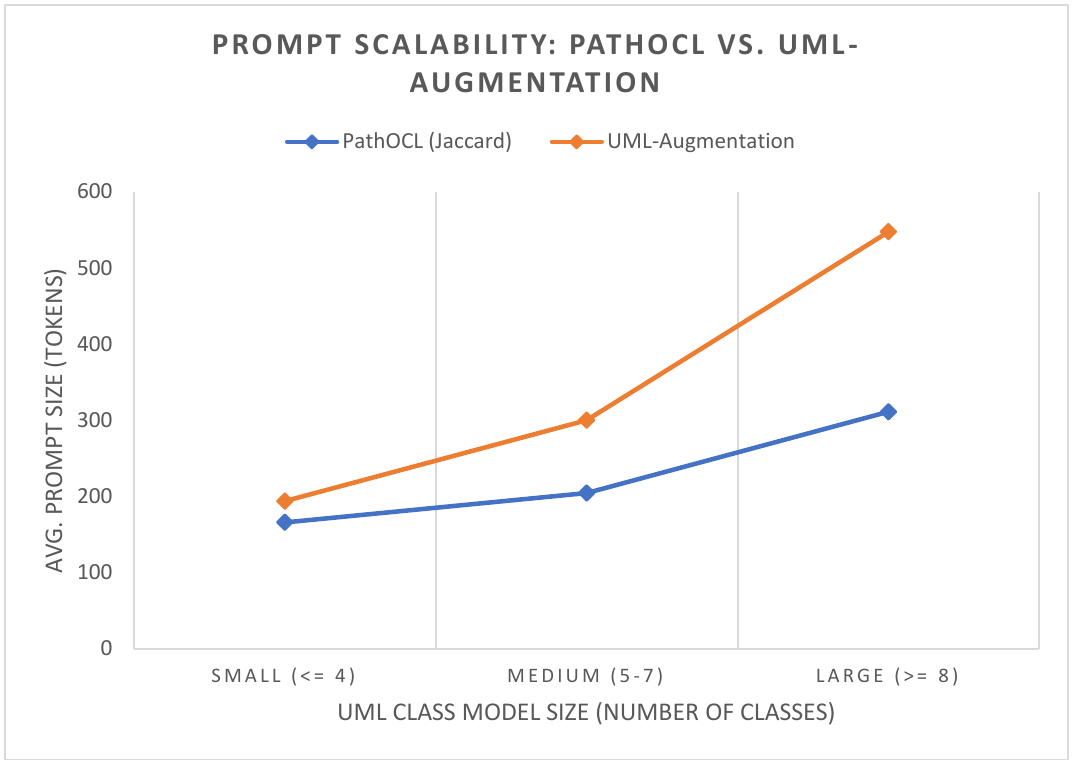}
    \caption{Trend analysis of the prompt size scalability for PathOCL (Jaccard) vs. UML-Augmentation with different UML class model sizes.}
    \label{fig:prompt-scalability}
\end{figure}

Figure \ref{fig:prompt-cost} shows that the top-1 prompt crafted using PathOCL is more cost-effective than UML-Augmentation. This is evident from the fact that the average prompt size in PathOCL is smaller, as shown in Figure \ref{fig:prompt-scalability}. However, PathOCL presents a trade-off: the inference costs increase substantially when achieving improved validity and correctness scores by covering top-k prompts with different simple paths, as confirmed by the results of our experiments in Table \ref{tab:top-k-results}. 

% Respective to our dataset
Now, to investigate the scalability of the prompts, we categorized the UML class models from our dataset into three groups based on the number of UML classes: small, medium, and large, as shown in Figure \ref{fig:prompt-scalability}. Our dataset contains 15 UML class models in total, with the smallest UML class model having 2 classes and the largest one containing 14 classes.

Based on the results shown in Figure \ref{fig:prompt-scalability}, we can observe that the average size of prompts crafted using both PathOCL and UML-Augmentation techniques is minimal for small-sized UML class models. However, the scaling trend becomes evident as the size of the UML class model grows larger. In addition, the difference becomes significant for larger UML class models, with the size of the prompt crafted using PathOCL reaching nearly half the size of the UML-Augmentation prompt. 

\section{Discussion}\label{SectionV}
Our proposed prompting technique, PathOCL, has shown a significant improvement when used with the GPT-4 model to express OCL constraints for UML class models. The improvement can be attributed to the core mechanism of PathOCL, which mimics the approach taken by experienced software practitioners when writing OCL constraints. Our research shows that by augmenting a selective subset of UML classes relevant to the English specification as context, the GPT-4 model is more likely to generate syntactically valid and semantically correct OCL constraints.

Also, we observed that by covering the simple paths, using the PathOCL technique could potentially generate unique and distinct OCL constraints for the same English specification. However, scalability limitations might occur as the complexity and size of the UML class models increase, deeming the brute force generation inefficient and raising concerns over the computational complexity of covering all simple paths. Therefore, further optimizations in the path coverage algorithm are necessary for the PathOCL technique to be applied in large-scale domain applications where computational resources are limited. 

PathOCL has demonstrated its advantage in chunking the UML class model to sub-models to reduce the context size of the prompt. This addresses a limitation faced by the UML-Augmentation technique when the size of the UML class model exceeds the context window of LLMs with smaller context, such as Llama 2, which has a limit of 4096 tokens \cite{touvron2023llama}. This impact is evident in Figure \ref{fig:prompt-scalability}, where the average prompts of larger UML class models (e.g., the Royal \& Loyal case) are almost half the size in comparison to UML-Augmentation prompts. This advantage is beneficial for large-sized domain models and is crucial for efficient on-device inference in resource-limited environments. 

However, the UML class models in our dataset, primarily compiled from educational resources, vary in size and may not accurately represent real-world applications. Therefore, an exhaustive dataset is required to ensure the applicability of PathOCL across diverse domain models.

Despite the marginal difference in the correctness of the generated OCL constraints between the two prompting techniques, this behavior can be expected from the GPT-4 model as a generalist model with the ability to achieve general-purpose language understanding and generation \cite{gpt-docs}. Therefore, the performance of GPT-4 is limited when it comes to software modeling tasks, such as understanding and expressing OCL constraints for UML class models. This limitation can serve as motivation for the software modeling community to adapt the intelligence of foundation models for domain modeling tasks \cite{Camara2023GPTModel, call_comm, Abukhalaf2023}.

\section{Related Work}\label{SectionVI}
Different NLP and model transformation techniques have been used to address the translation of natural language (i.e., English) specifications to OCL constraints. One approach, known as COPACABANA \cite{Wahler2008UsingPT}, is a semi-automatic pattern-based approach that requires human intervention during the translation process. Another approach is NL2OCLviaSBVR \cite{bajwa2010ocl}, an MDA-based tool that automatically transforms English specifications to an intermediate representation in Semantics of Business Vocabulary and Rules (SBVR), and then to OCL constraints. Salemi et al. \cite{Salemi2016OCL} proposed another approach called English2OCL (En2OCL). Their approach takes the English specification and the UML class model as input to generate the required OCL constraint. It also includes the definition of mapping rules to analyze the English specification and extract the UML elements that can be mapped to their equivalent OCL expression. 

Recent studies explored the reliability of LLMs in the model-driven development domain. In our previous study \cite{Abukhalaf2023}, we evaluated Codex in generating OCL constraints from English specifications using different prompt templates and techniques. We augmented the entire UML class model in the prompt and assessed both zero- and few-shot prompting settings. The results demonstrated that providing the UML class model as context significantly improved the reliability of the OCL constraints. Another study by Camara et al. \cite{Camara2023GPTModel} explored the capabilities of ChatGPT in supporting domain modeling tasks. Their findings demonstrated that ChatGPT performed well in dealing with small domain models, although it had some limitations in basic modeling concepts. Additionally, they also observed remarkable performance in generating OCL constraints.

\section{Threats to Validity}\label{SectionVII}
\textbf{Large language model.}
The primary concern revolves around any potential biases within the GPT-4 model itself. As the training dataset of GPT-4 is not publicly disclosed, there may be a risk of bias. The dataset we used in our study was curated from educational resources that are publicly available on the internet and GitHub. Hence, there is a possibility that UML class models, English specifications, and their OCL constraints could be present in the training dataset. In addition, OpenAI hosts the GPT-4 model and offers it as an API service. However, it is important to note that the performance may vary and results might not be reproduced identically, even with a zero temperature setting. Therefore, consistent behavior cannot always be guaranteed.

\textbf{Quality of the dataset.} 
The UML class models in our dataset are sourced from educational resources. As such, it is important to acknowledge that the English specifications and their OCL constraints may not be exhaustive. Therefore, our findings may not entirely reflect real-world applications. A more diverse dataset would lead to a robust and comprehensive evaluation of PathOCL and other prompting techniques.

\textbf{USE modeling tool.} 
The USE modeling tool, as stated on its GitHub repository, is primarily designed as a research prototype \cite{GOGOLLA200727USE}. As a result, it may not be fully developed, thoroughly tested, or stable. This could lead to unexpected behaviors or limitations that may impact our study results when evaluating the OCL constraints.

\section{Conclusion}\label{SectionVIII}
In this study, we present PathOCL, a path-based prompting technique that combines natural language processing with simple path coverage to selectively augment a subset of UML classes relevant to the specified requirements. Our research demonstrates that PathOCL improves the validity and correctness of the OCL constraints generated by the GPT-4 model compared to the UML-Augmentation technique. 

Overall, the PathOCL technique significantly improved the validity score. However, there was a slight increase in the number of correctly generated constraints, suggesting a minimal impact on semantic correctness. This aligns with the general-purpose nature of GPT-4, which faces limitations in specialized domains such as software modeling and OCL constraints.

Additionally, we evaluated the inference costs and the scalability of prompts crafted when using the PathOCL technique. The current implementation of PathOCL has shown its effectiveness in reducing the context size of prompts by nearly half when scaling UML class models. However, testing prompts with distinct simple paths is necessary to achieve improved scores over the UML-Augmentation technique. This represents a trade-off decision when adopting the PathOCL technique.

In conclusion, PathOCL emerges as a promising technique that can improve the effectiveness of LLMs, such as GPT-4, in generating reliable OCL constraints. Although it signifies progress in automated OCL generation, further optimization of the PathOCL approach and adaptation of foundation models for software modeling tasks could lead to more widespread use of the OCL language.

%Bibliography
\bibliographystyle{unsrt}  
\bibliography{references}

\end{document}